# Macroscopic self-reorientation of interacting two-dimensional crystals

C. R. Woods[1], F. Withers[1], M. J. Zhu[1], Y. Cao[1], G. Yu[1], A. Kozikov[1], M. Ben Shalom[1], S. V. Morozov[1,2,3], M. M. van Wijk[4], A. Fasolino[4], M. I. Katsnelson[4], K. Watanabe[5], T. Taniguchi[5], A. K. Geim[6], A. Mishchenko[7], K. S. Novoselov[1,7]

[1]School of Physics and Astronomy, University of Manchester, Oxford Road, Manchester, M13 9PL, UK
[2]Institute of Microelectronics Technology and High Purity Materials RAS, Chernogolovka, 142432, Russia
[3]National University of Science and Technology "MISiS", Moscow, 119049, Russia
[4]Radboud University, Institute for Molecules and Materials, Heyendaalseweg 135, 6525 AJ Nijmegen, the Netherlands
[5]National Institute for Materials Science, 1-1 Namiki, Tsukuba 305-0044, Japan
[6]Centre for Mesoscience and Nanotechnology, University of Manchester, Oxford Road, Manchester, M13 9PL, UK
[7]National Graphene Institute, University of Manchester, Oxford Road, Manchester, M13 9PL, UK

*Microelectromechanical systems, which can be moved or rotated with nanometre precision, already find applications in such fields like radio-frequency electronics, micro-attenuators, sensors and many others. Especially interesting are those which allow fine control over the motion on atomic scale due to self-alignment mechanisms and forces acting on the atomic level. Such machines can produce well-controlled movements as a reaction to small changes of the external parameters. Here we demonstrate that, for the system of graphene on hexagonal boron nitride, the interplay between the van der Waals and elastic energies results in graphene mechanically self-rotating towards the hexagonal boron nitride crystallographic directions. Such rotation is macroscopic (for graphene flakes of tens of micrometres the tangential movement can be on hundreds of nanometres) and can be used for reproducible manufacturing of aligned van der Waals heterostructures.*

In many layered crystals it is the van der Waals interaction which is responsible for perfect stacking of individual layers. Once such perfect stacking is lost (for instance through a rotational fault), the van der Waals interaction tends to restore the perfect stacking – the effect known as self-rotation. This effect has been seen for nanometre sized graphene flakes when driven by an atomic force microscopy (AFM) tip on the surface of graphite[1]. However, up to now, such phenomena has not been observed at micrometre or larger sizes, apart for the cases when such self-rotation was driven by surface free energy in displaced graphite mesa structures[2, 3].

One of the reasons why self-rotation is hard to observe in homogeneous systems (where the two surfaces are represented by the same crystals) is because both the self-rotating forces (which try to return the crystals to perfect stacking), and the friction forces are essentially determined by the same van der Waals potential. So, even when close to the perfect commensurate state (where the self-retracting forces should be the strongest) the van der Waals potential would exhibit a number of local potential minima (which correspond to strong friction), where the system may get localised.

The situation is very different when the two crystals are not identical (for instance, have different lattice constants). In this case the local minima in the van der Waals potential are not expected to play such a significant role, because of strong incommensurability. Additionally, if at least one of the crystals has the freedom to relax elastically, the van der Waals potential starts to compete with elastic energy, forming more complex potential landscape. Thus, it is interesting to investigate if the self-rotation can be achieved in such heterogeneous structures.

Such interfaces can be created by stacking several 2D atomic crystals into van der Waals heterostructures[4-6], with one of the most interesting systems being graphene on hexagonal boron nitride (hBN)[7], since the lattice constants of the two crystals are different only by 1.8%. It has been shown that graphene on hBN has an observable moiré pattern, whose period depends on the misorientation angle[8, 9]. Due to the difference in the interatomic distances for the two crystals, the maximum moiré period (of ~14nm) is achieved when the crystallographic lattices are perfectly aligned. At small deviations from the alignment, graphene on hBN undergoes an incommensurate to commensurate transition[10]. In the commensurate state, graphene splits into domains (where its lattice is stretched to gain in van der Waals interaction energy with hBN) separated by sharp domain walls (where graphene lattice is relaxed)[10, 11]. Within the domain the stretching is gradual, ranging[11] from more than 1%, down to 0%. Thus, the average stretching of graphene is quite small (well below 1.8%), resulting in only a small lost in the elastic energy, which is compensated by the gain in the van der Waals energy.

Such stretching of graphene, even so being small, leads to global breaking of the sublattice symmetry[12-14]. Thus, the possibility to align graphene and hBN is extremely important, and already led to the observation of a number of exciting physical phenomena, such as Hofstadter butterfly[15-17] and topological currents[18]. Furthermore, the concept of self-alignment could be extended to other interfaces and utilised for the formation of novel devices[19-26] which rely on such aligned crystals (for example, resonant tunnelling diodes[27]).

Typically the commensurate state is identified by a small (of the order of 0.1) ratio between the width of the domain walls ($\delta$) and the moiré period ($L$), whereas $\delta/L \approx 0.5$ in the incommensurate phase.

In the following, we demonstrate that, despite the strong competition between the elastic and van der Waals energies, graphene can reorient itself on top of hBN towards a commensurate state (where the crystallographic axis of the two crystals are aligned better than ~ 0.7°).

Graphene flakes, studied in this work, were transferred onto hBN by the dry transfer method[28, 29], to produce a clean interface, Fig. 1A. During the transfer procedure we ensure (by direct optical observation of the crystallographic facets in the transfer set-up) that the crystallographic directions of graphene and hBN are misoriented by $\theta$=1-2 degrees. We further confirmed the misorientation angle by measuring the period of the moiré pattern in scanning probe experiments[8-10], Fig. 1C, as well as by the width of Raman 2D peak (Fig. 2A) which can be related to the period of the moiré superstructure[30] and the misorientation angle[30], Fig. 2B. Moiré patterns can be observed in various channels in AFM, including topography, friction, *etc*., as well as in scanning tunnelling experiments[8, 9] and conductive AFM[15]. Here we mainly used PeakForce Tapping mode[31] and evaluated the point Young's modulus channel with a typical resolution better than 2 nm.

Fig. 1A shows an optical image of one of our graphene on hBN structures (another example is given in Supplementary Note 1). Originally it has been aligned by $\theta \approx 1.0°$ with respect to the hBN flake, as confirmed by AFM (Fig. 1B) and Raman (Fig. 2A, B). We would like to note that, even prior to annealing, this flake approaches the commensurate state ($\delta/L$=0.35, Fig. 1D). The sample was annealed at 200°C for four hours in forming gas (90% Ar + 10% H$_2$). After annealing $L$ increases by 15% (from 10nm to 11.5nm, Fig. 1B), which indicates greater alignment (misalignment angle $\theta \approx 0.7°$). Importantly, $\delta/L$=0.20 after annealing, which demonstrates an increased level of commensuration (also confirmed by Raman, Fig. 2C, D). The alignment is uniform across the flake, which could be seen from the Raman signal (Fig. 2B, D) or from the observation of the uniform moiré period by AFM measurements in different parts of the sample (See Supplementary Note 2, and similar data for another sample in Supplementary Note 1).

We would like to stress that neither formation of creases nor strain accumulation have been observed after the annealing (as follows from our AFM and Raman measurements, respectively). To achieve such uniform alignment the graphene flake should have uniformly rotated by $\Delta\theta$=0.3$^o$. It means that some parts of the flakes should have moved by $d=\Delta\theta l\approx$0.15μm (here $l\sim$30μm is the characteristic size of the flake). This is a significant macroscopic movement which can be used to drive certain nanomachines (such a macroscopic motion is demonstrated in Supplementary Note 1, as well as has recently been seen by other groups as well[32]).

What pushes such macroscopic movement is the gradient in the van der Waals forces. To analyse their role, we compare, in Fig. 3, the interlayer van der Waals energy to the elastic intralayer contribution to the total energy after energy minimization for different alignments, relative to the values at 0$^o$. The total energy does not vary up to 0.7$^o$ degrees, after which the interlayer energy interaction increases while the intralayer energy decreases, resulting in an increase of total energy. As all values are obtained from energy minimization from a given angle, this figure does not give information about the barriers between different angles.

This picture fits remarkably well with our experimental observation. Our graphene flakes rotated to within 0.7$^o$ to the crystallographic orientation of hBN, which corresponds nicely to the plateau in van der Waals energy misalignment dependence for $\theta$≲0.7$^o$. Still, we note, that in many previous experiments[8-10, 15-17] the graphene flakes exhibit much better alignment than 0.7$^o$, which we would like to also attribute to the self-rotation mechanism.

We would like to stress that not all the flakes become aligned after annealing. We had a number of flakes which do not self-align. At the same time, those which do not undergo the self-rotation would typically form one dimensional network of wrinkles, Fig. 4B, similar to that reported previously[33] (although in that case such wrinkles are formed upon cooling). Similar to the moiré pattern, the wrinkles could be observed in several AFM modes, though they are most clearly visible in the local Young modulus and height channels (See Supplementary Note 3 for more examples). The fact that they are readily observable in the Young's modulus channel, suggests strain accumulation around the wrinkles, which is also confirmed by an increase in the FWHM of the Raman 2D peak, Fig. 4E (in this case the broadening is uniaxial, which reflects the fact that wrinkles predominantly create strain only in one direction, see Supplementary Note 4). Fig 4(A) and Fig 4(B) show the contrasting images in Young's modulus of the moiré pattern before and after annealing to high temperature respectively. The one-dimensional network of wrinkles is clearly visible on the sample after annealing to be superimposed on the moiré structure, Fig. 4B. At the same time, the period of the moiré structure has not changed. We would also like to suggest that the wrinkles are most likely linked to the moiré structure, as seen from the orientation and the position of the peaks in the Fourier transform patterns.

The proposed mechanism for their formation is the following. At high temperature, due to the difference in the thermal expansion coefficients (TECs) between the hBN and graphene, the lattice mismatch increases, favouring the incommensurate phase. Upon cooling, the same difference in TEC acts as a compression for graphene, possibly leading to wrinkles, Fig. 4F. Also, the reconstructed moiré pattern recovers upon cooling, making the two structures (wrinkles and the reconstructed moiré pattern) to coexist, Fig. 4G. However, as both the domain walls of the reconstructed moiré pattern and the wrinkles carry strain field, it becomes energetically favourable to make the two commensurate, overlapping the wrinkles and the domain walls, Fig. 4H. Furthermore, the wrinkles can undergo further reconstruction within the domains of stretched graphene themselves, Fig. 4I. In this model wrinkles should carry additional strain, which indeed has been observed by strong broadening of the Raman 2D peak, Fig 4E. Such contribution to the strain energy changes the potential landscape and, as it turns out, prevents the self-alignment process.

Finally, the presence of contamination bubbles and creases seems to prevent the possibility of self-alignment. Self-alignment was not observed in any sample with more than a few bubbles. The detrimental influence of the bubbles can be twofold: it reduces the interaction area between graphene and hBN, making the van der Waals potential landscape shallower; also, the contamination concentrated in such bubbles[29] can act as pinning centres, preventing any macroscopic movements of graphene.

Our observation opens a new direction in the physics and applications of van der Waals heterostructures – self aligned stacks. Already now this effect is being used to produce graphene aligned on hBN for transport experiments (such as the observation of Hofstadter effect[15-17], topological currents due to Berry curvature[18], *etc.*). In such devices the self-rotation could be in principle observed directly as shifting of the secondary Dirac point and the associated with it resistance peak (see Supplementary Note 5). We also expect that such self-rotation is not unique to graphene/hBN stacks and that other layered materials should exhibit similar behaviour. For instance, in Supplementary Note 6 we present an example of the observation of self-rotation in graphene/graphene stack being seen via direct measurements of the electronic density of states in tunnelling experiments. Furthermore, one can utilise the mechanical motion of the crystals to produce nanomechanical devices. It is still unclear to what extent the surface reconstruction of the crystal influences the van der Waals potential – a subject still to be explored further both through experimental and theoretical investigations.

Our samples were produced by the dry ('stamp') transfer technique described in detail previously[28,29,34]. In brief, the method involves using a double polymer layer to identify and isolate graphene flakes on a membrane, before bringing the graphene into contact with the hBN. Importantly, this method doesn't require the use of any solvents, which minimises the contamination.

AFM measurements were performed on a Bruker FastScan atomic force microscope, in the PeakForce[31] feedback mode, which allows the extraction and analysis of individual force curves for each pixel at regular scanning speeds (0.5 – 4 Hz). Typically, fast and large area scans are used to determine the period, whilst slower and smaller area scans are used to calculate the ratio $\delta/L$. Raman spectroscopy measurements were taken with the Witec confocal Raman spectrometer with a wavelength of 514 nm and 1mW power.

For the calculation of the interaction energies, we constructed a model of graphene on h-BN with their crystallographic axis rotated with respect to each other. The h-BN is kept fixed to mimic a bulk substrate. Note that a different supercell has to be constructed at each misorientation angle (see Supplementary Note 7). The size of the supercells with periodic boundary conditions demands the use of an empirical potential. The graphene atoms interact through the reactive empirical bond order potential REBO[35], as implemented in the molecular dynamics code LAMMPS[36]. This potential is widely used in simulations of carbon materials in view of its excellent description of structure and elastic properties of all carbon allotropes. As no potential for graphene/hBN interaction is currently available, the interlayer interaction is assumed to be of the form of a registry-dependent potential for interlayer interactions in graphene[37], without the correction for bending. We scale this potential to the lattice constant of h-BN and use different scaling factors for C-B and C-N interactions as was done in Ref.[11] because this leads to a good agreement with experimental results[10] and ab-initio calculations[38, 39]. We have further refined this approach[13] leading to a choice of the B-C interaction of 60% of the C-C value while the N-C interaction is set to 200% of the C-C value in the original form[37]. We minimize the total potential energy by relaxing the graphene layer by means of FIRE[40], a damped dynamics algorithm. For samples close to alignment this leads to significant changes in bond length along the moiré pattern.

## Acknowledgements

This work was supported by The Royal Society, U.S. Army, European Research Council, EC-FET European Graphene Flagship, Engineering and Physical Sciences Research Council (UK), U.S. Office of Naval Research, U.S. Air Force Office of Scientific Research, FOM (The Netherlands). S.V.M. is supported by NUST "MISiS" (grant K1-2015-046) and RFBR (14-02-00792). M. J. Zhu acknowledges the National University of Defense Technology (China) overseas PhD student scholarship.

## Author Contribution

C.R.W. produced experimental devices, measured device characteristics, analysed experimental data, participated in discussions, contributed to writing the manuscript; F.W., M.B.S. and Y.C. produced experimental devices; M.J.Z., S.V.M. and G.Y performed transport measurements; A.K contributed to Raman studies; M.M.v.W, A.F., M.I.K. provided theoretical support; K.W. and T.T. provided hBN. A.K.G. analysed experimental data, participated in discussions, contributed to writing the manuscript; K.S.N. initiated the project, analysed experimental data, participated in discussions, contributed to writing the manuscript; A. M. analysed experimental data, participated in discussions, contributed to writing the manuscript.

## Additional information

Correspondence and requests for materials should be addressed to K.S.N. (kostya@manchester.ac.uk).

# Figures and Legends

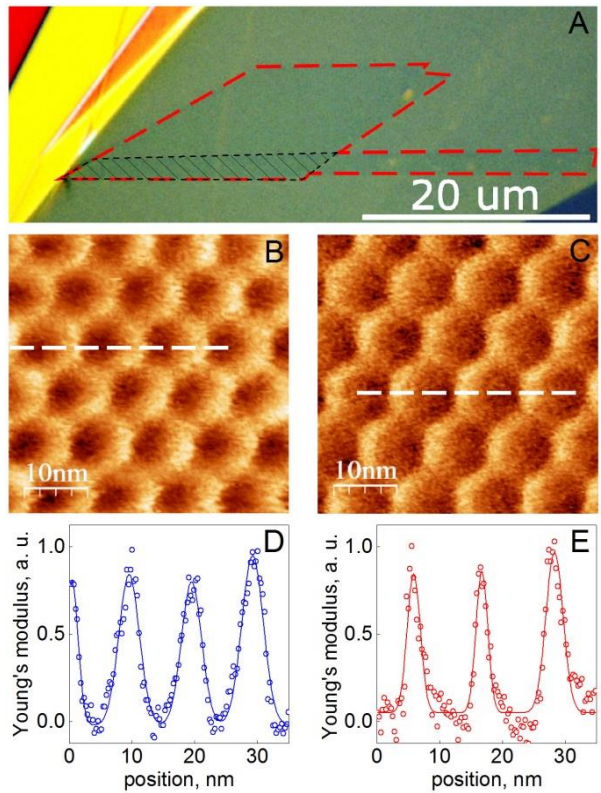

*Fig. 1 (A) Optical microscopy image of the flake, demonstrating a very clean interface (bubble free) between graphene and hBN. Different colours correspond to different thicknesses of hBN. Graphene is practically invisible and is marked by red dashed line. The hatched area is bilayer graphene. (B) and (C) Young modulus distributions obtained in PeakForce Tapping mode of the moiré superlattice before (B) and after (C) self-alignment. (D) and (E) line profiles across the respective Young modulus distribution images, which indicates the smaller width of the Young's modulus peaks in the annealed (self-rotated) sample. Symbols – experimental data, solid curves – fitted peaks.*

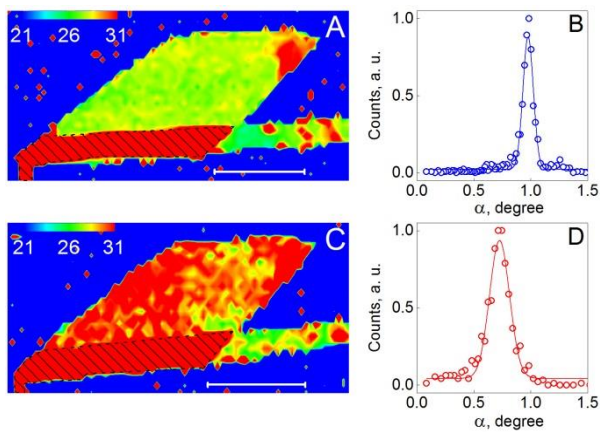

*Fig. 2 (A) and (C) Maps of the FWHM of Raman 2D peak before and after annealing respectively. (B) and (D) histograms alignment angles, as recalculated from (A) and (C) respectively.*

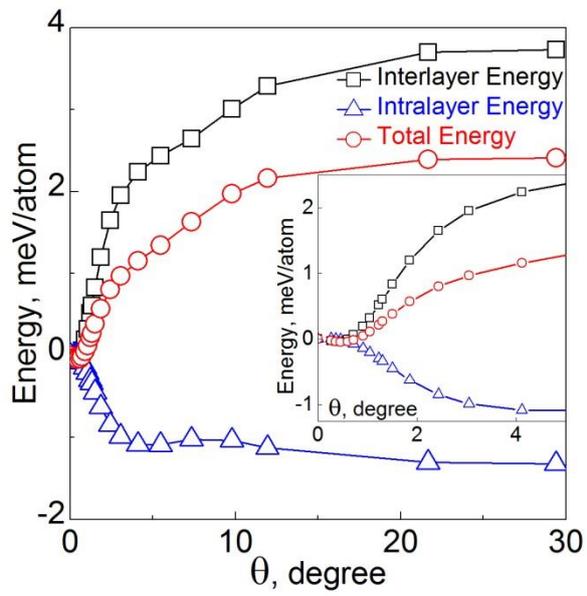

*Fig. 3 – Total energy (red) contributions from intralayer (elastic changes/blue) and interlayer (adhesive/black) interactions, as a function of alignment angle, relative to the value at θ=0. Points are calculated by minimising energy for a given angle. Inset: the same curves for low misalignment angles.*

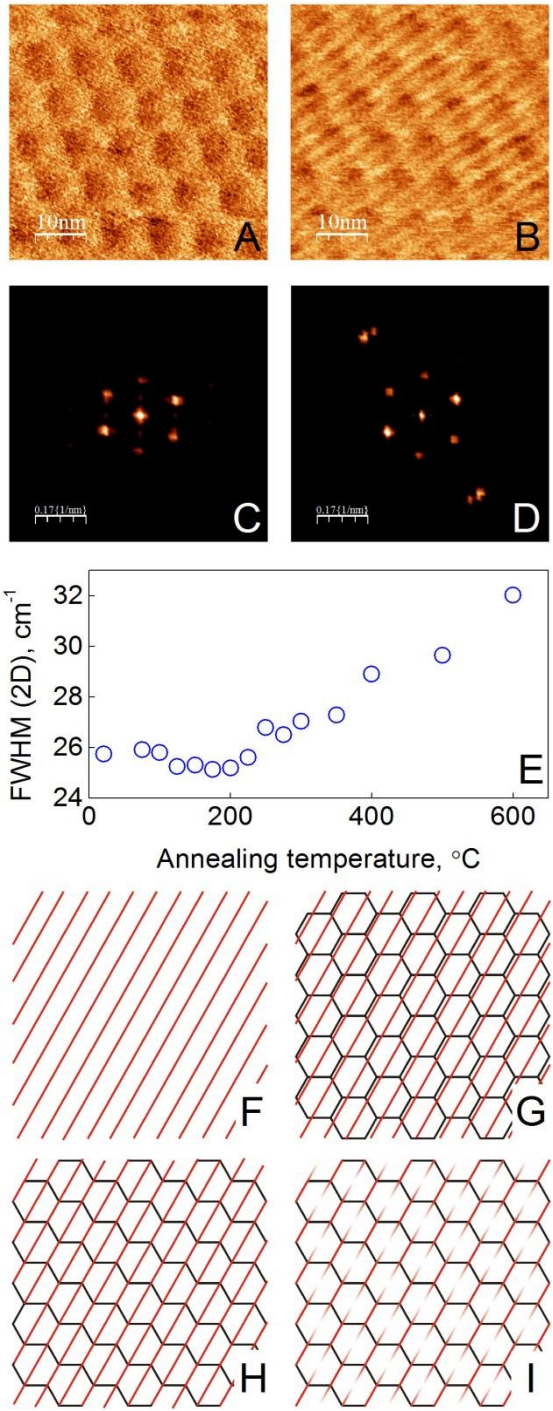

*Fig. 4 – (A) and (B) AFM images of the moiré superstructure in a sample which did not rotate before and after annealing to 600°C respectively. (C) and (D) are the Fourier transformations of (A) and (B) respectively. (E) Width of the 2D peak in the Raman scattering spectrum as a function of annealing temperature, the increase is linked to the formation of 1D wrinkles. (F)-(I) Proposed structure of the superposition between the moiré pattern and the 1D wrinkles. At high temperatures 1D wrinkles are formed due to difference in thermal expansion coefficients of graphene and hBN (F). Upon cooling the moire structure appears which coexists with the wrinkle (G). It is more energetically favourable, however, for the 1D wrinkles to coincide with the domain walls of the moiré structure (H). Part of the wrinkle can be flattened due to commensurate-incommensurate transition (I).*

# Supplementary Information

# Supplementary Figures

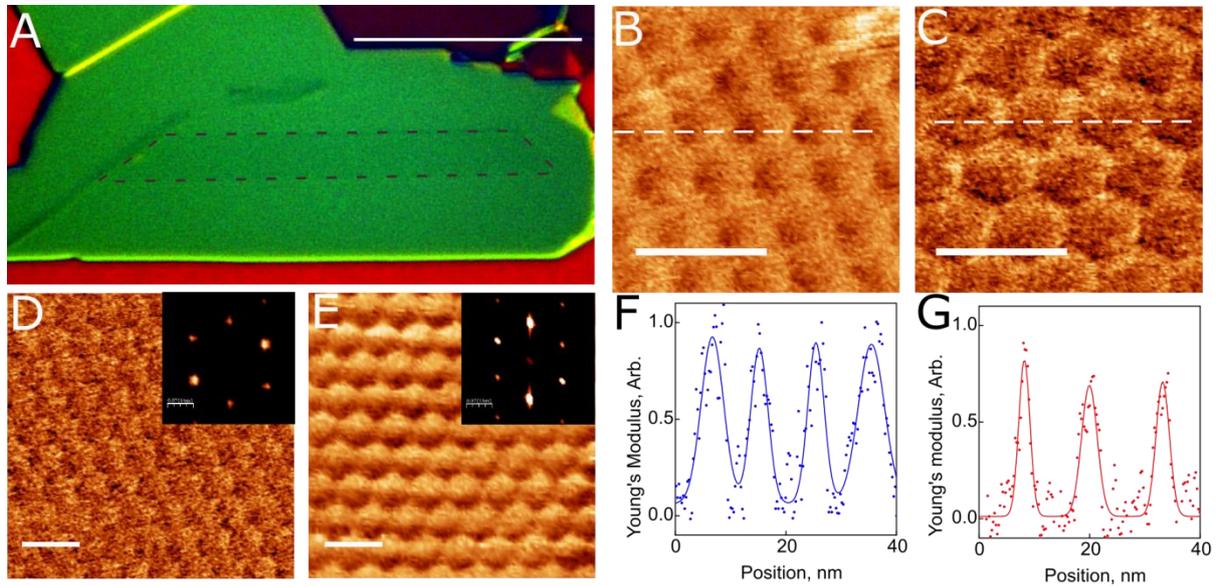

***Supplementary Figure 1. Optical and Atomic force microscopy data for a self-rotating flake.*** *(A) optical microscopy image of the flake showing the clean interface (scale bar is 20 µm), (B) and (C) Young's modulus images of the moiré superlattice before and after self-alignment respectively (Scale from black to white: 60 to 75 MPa, and 45 to 60 MPa), and (D) and (E) larger scale Young's modulus images of the moiré patterns used to extract the periodicity for before and after annealing respectively (Scale from black to white: 60 to 84 MPa, and 40 to 62 MPa). The scale bar in (B), (C), (D), and (E) is 20 nm. (F) and (G) are line profiles of the dashed lines in the Young's modulus maps (B) and (C) respectively. Images (B), (C), (D), and (E) were taken centrally in the right hand region of the flake.*

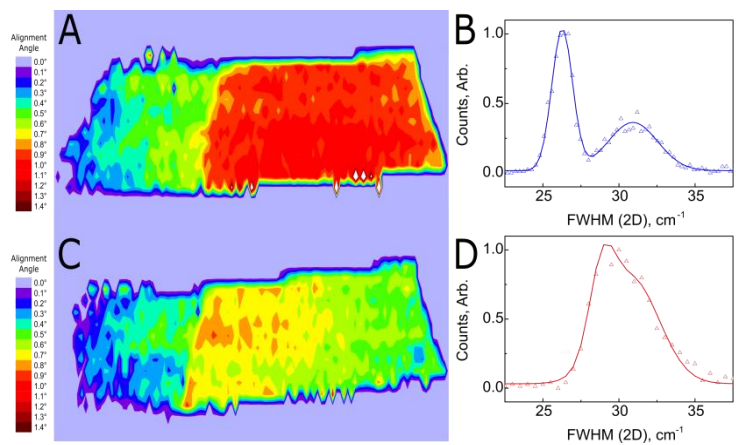

***Supplementary Figure 2. Raman spectroscopy data for the self-rotating flake.*** *(A) and (C) Alignment maps of the flake before and after annealing respectively, taken from the broadening of the 2D peak in the Raman spectrum. (B) and (D) histograms of (A) and (C) respectively.*

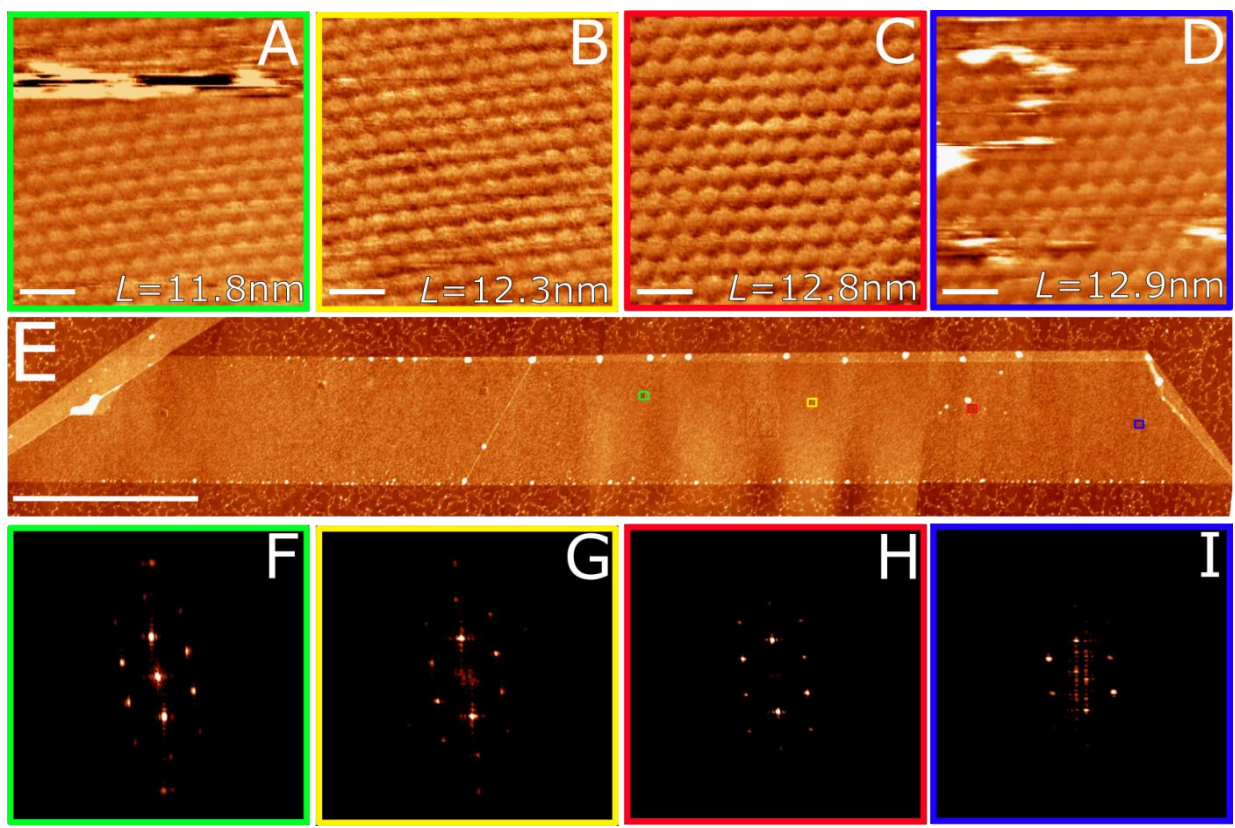

***Supplementary Figure 3. A comparison of the moiré pattern at different points on the flake, by AFM.*** *(A) (Green), (B) (Yellow), (C) (Red), and (D) (Blue) Young's modulus images of the moiré pattern at 5 um intervals along the flakes length (to the right of the fold). These show the gradual change in moiré periodicity. The scale bar in (A), (B), (C), and (D) is 25 nm. Some contamination on the flake surface remains, which is visible in images (A) and (D). (E) an AFM image of the entire flake, with the positions of (A), (B), (C), and (D) indicated by their colours (Also, in order left to right). Also present in the image is the fold which creates a discrete change in relative orientation angles. The scale bar for (E) is 5 μm. (F), (G), (H), and (I) show the Fourier transform of images (A), (B), (C), and (D) respectively. Scales, from black to white, in (A), (B), (C) and (D) are; 119 to 145 MPa, 120 to 140 MPa, 120 to 140 MPa, 110 to 145 MPa.*

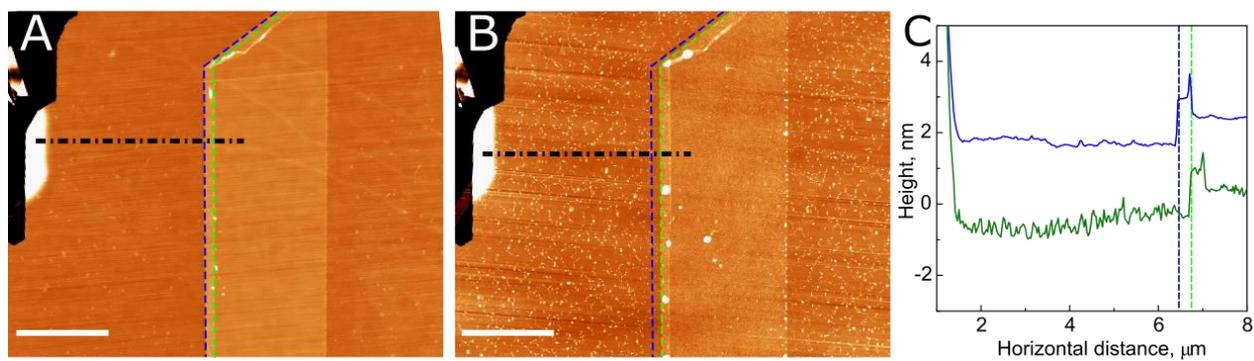

***Supplementary Figure 4. AFM data demonstrating the lateral displacement of the flake.*** *AFM height images of the graphene flake (A) before, and (B) after annealing, with the edge of the flake before (Blue) and after (Green) marked. The scale bar in (A) and (B) is 3 µm. (C) Profiles (averaged over 10 lines) of the black dashed from (A) and (B) line indicating the change in position of the step by 300 ± 90 nm. This step corresponds to the start of the flake.*

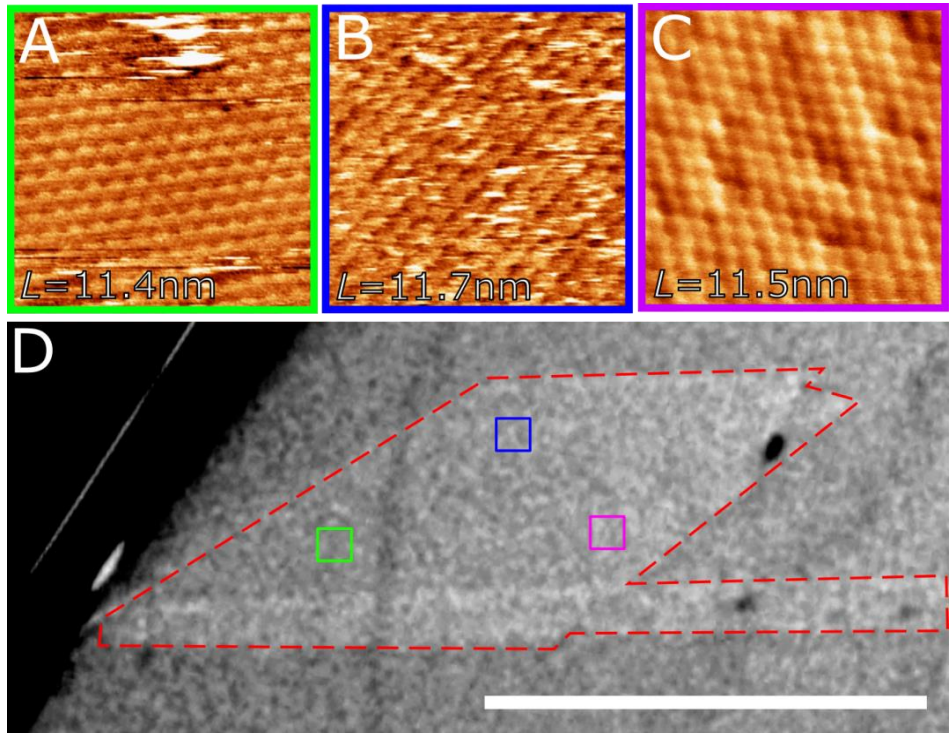

***Supplementary Figure 5. AFM data revealing the homogeneity of the moiré pattern.*** *(A) (Green), (B) (Blue), and (C) (Purple), Young's modulus maps of the moiré pattern at various points on the flake. Each image is 150x150nm. Some contamination remains in the images. Scale from black to white for (A), (B) and (C) is; 31 to 42 MPa, 30 to 45 MPa, and 30 to 45 MPa. (D) Optical image of the flake with approximate regions on the flake for the scans of (A), (B), and (C). The scale bar in (D) is 20 µm.*

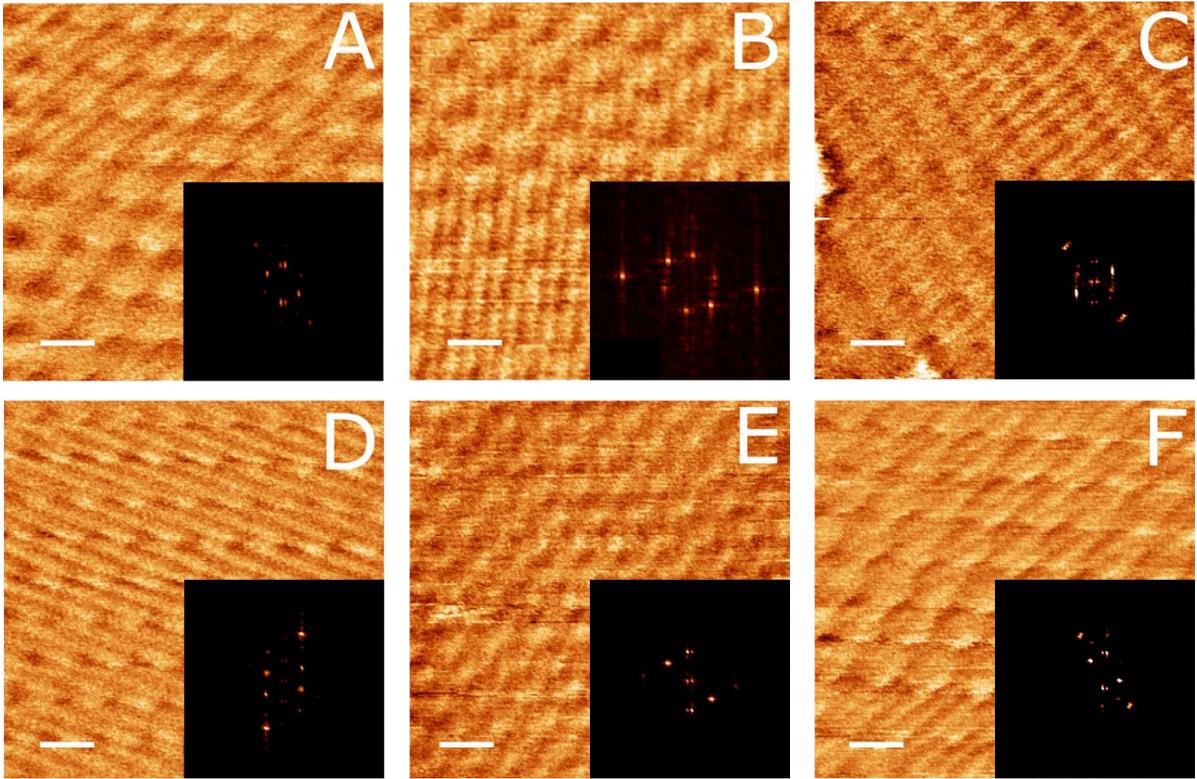

***Supplementary Figure 6. 1D wrinkling in various samples.*** *(A), (B), (C), (D), (E), and (F) Young's modulus maps of the moiré superlattice with the wrinkles also present. In each image the scale bar is 10 nm. Inset to each image is the Fourier transform of the image, which shows the two points associated with the 1D wrinkles. Scales for (A), (B), (C), (D), (E), and (F) are (from black to white); 13 to 19 MPa, 20 to 32 MPa, 100 to 130 MPa, 35 to 49 MPa, 8 to 12 MPa, and 29 to 40 MPa.*

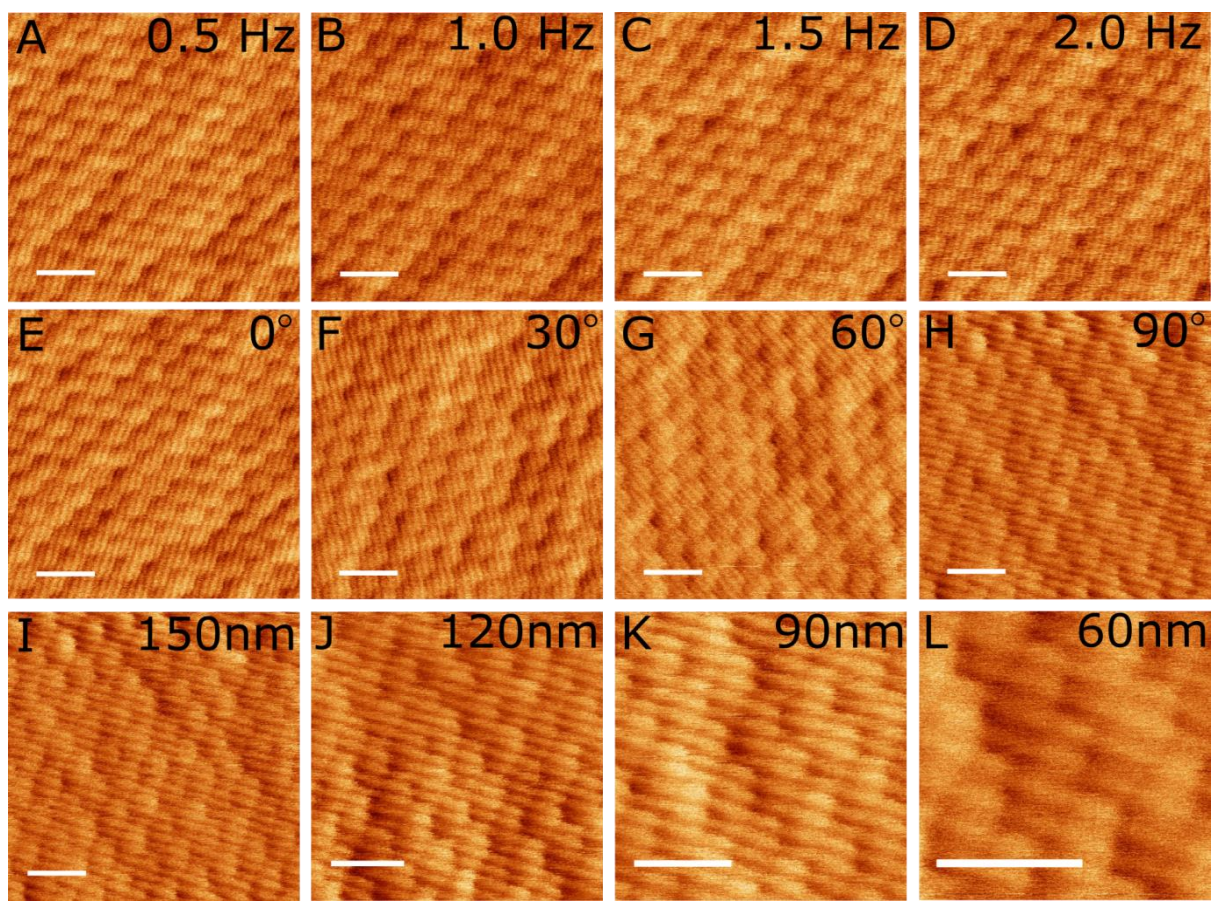

***Supplementary Figure 7. AFM data demonstrating that the 1D wrinkling is not an scanning artefact.*** *(A), (B), (C), and (D) Young's modulus maps showing the periodicity and direction of the 1D wrinkling (and graphene-hBN superlattice) at a scan rate of 0.5 Hz, 1.0 Hz, 1.5 Hz, and 2 Hz respectively. (E), (F), (G), and (H) Young's modulus maps showing that the orientation change of the moiré superlattice and 1D wrinkles for a scan angle of 0°, 30°, 60°, and 90° respectively. (I), (J), (K), and (L) Young's modulus maps showing the moiré superlattice and 1D wrinkles at 150 nm, 120 nm, 90 nm, and 60 nm scan sizes. In each row only one scanning parameter was changed. Some anisotropic lateral drift is present in the images, however it is not significant. The scale bar in each image is 30 nm.*

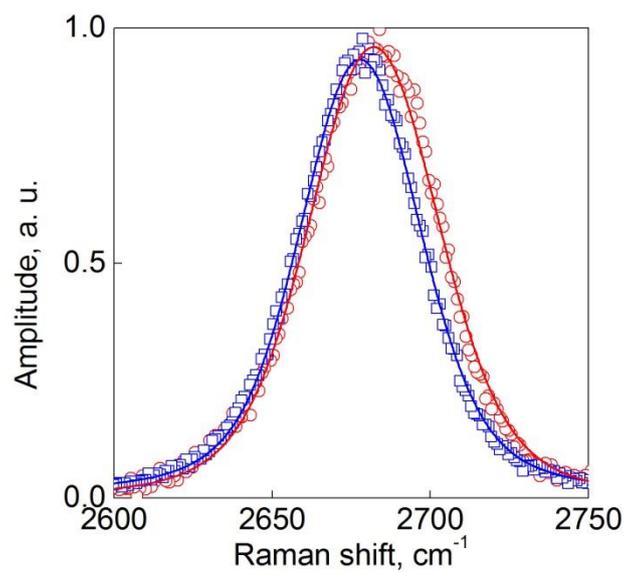

***Supplementary Figure 8. Polarised Raman spectroscopy data for 1D wrinkles.*** *Raman 2D peak for a sample with uniaxial wrinkling (symbols – experimental data, curves - fitting). Red circles and curve – for linear polarization of incoming light perpendicular to the wrinkles, blue squares and curve – for linear polarization of incoming light along the wrinkles.*

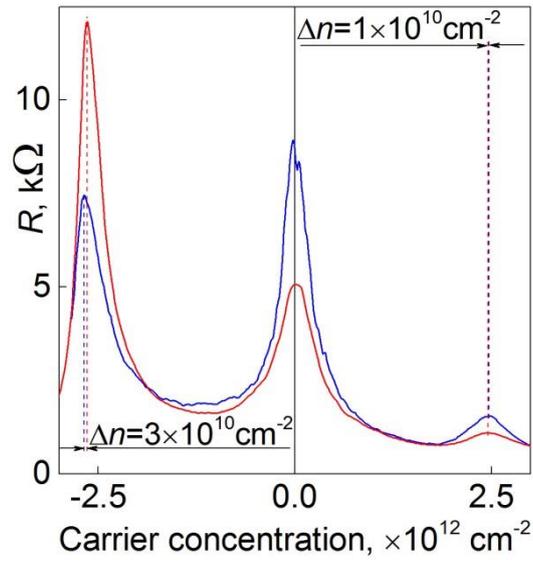

***Supplementary Figure 9. Transport data for a self rotating graphene on hBN sample.*** *Resistance as a function of carrier concentration (measured at 4K) for a graphene on hBN device annealed at 175 ° (blue curve) and 200 ° (red curve). The shift in position of the secondary Dirac points indicate self-rotation towards more aligned state.*

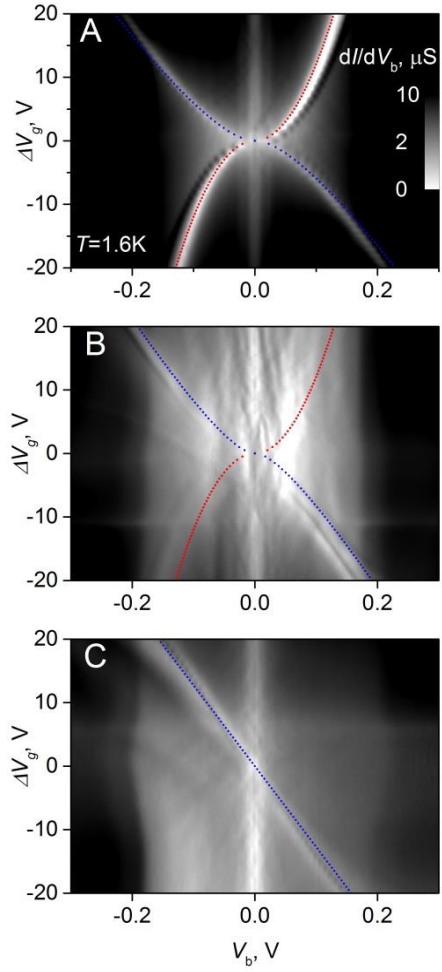

***Supplementary Figure 10. Differential conductivity (in logarithmic scale) as a function of the bias and gate voltages for Si/SiO$_2$/graphene/hBN/graphene structure.*** *(A), Si/SiO$_2$/graphene/hBN/graphene/graphene structure (B) and annealed Si/SiO$_2$/graphene/hBN/graphene/graphene structure (C).The blue (red) curves are the calculated positions of the event when the Fermi level in the bottom (top) graphene electrode passes through the neutrality point.*

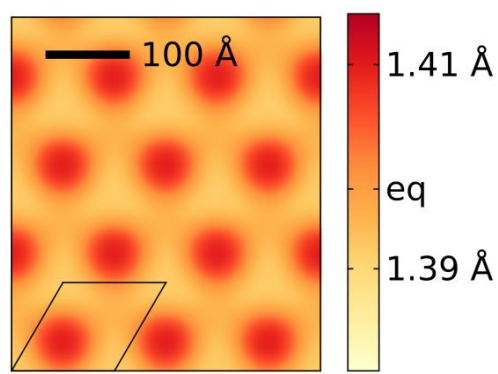

***Supplementary Figure 11. Distribution of bond lengths for the relaxed graphene on hBN.*** *Calculated for θ=0. The equilibrium value given by the REBO potential as implemented in LAMPPS, of isolated graphene, is 1.3978 Å.*

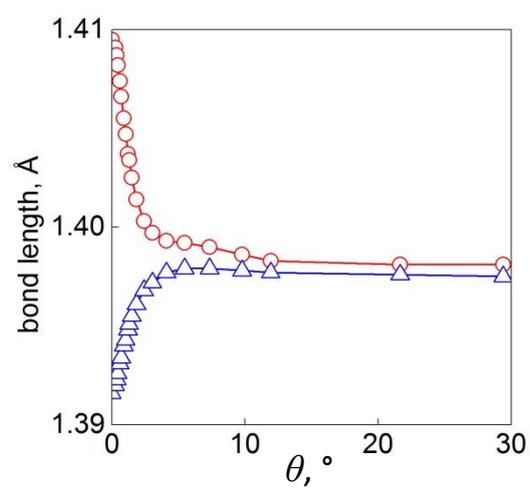

***Supplementary Figure 12. Theoretical bond length as a function of misorientation angle.*** *Maximum (red) and minimum (blue) values of the bond lengths as a function of the misalignment angle $\theta$.*

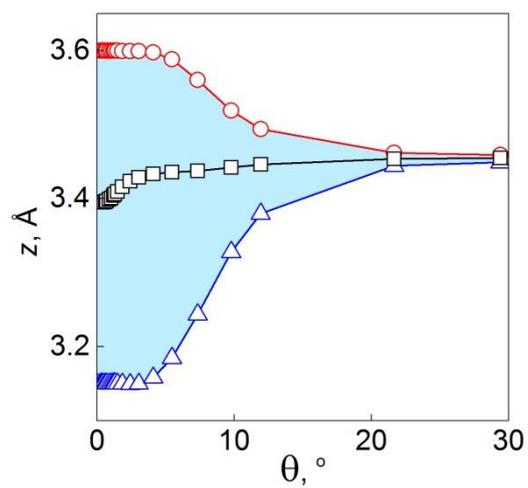

***Supplementary Figure 13. Vertical separation between graphene and hBN.*** *The three lines show the maximum (red), average over all atoms (black), and minimum (blue) values of atomic distances from the substrate as a function of the misalignment angle θ.*

# Supplementary Tables

| *n* | *m* | *q* | *θ* |
|---|---|---|---|
| 55 | 0 | 56 | 0.0 |
| 192 | 1 | 196 | 0.26 |
| 137 | 1 | 140 | 0.36 |
| 219 | 2 | 224 | 0.45 |
| 164 | 2 | 168 | 0.6 |
| 272 | 4 | 279 | 0.72 |
| 217 | 4 | 223 | 0.91 |
| 378 | 8 | 389 | 1.04 |
| 161 | 4 | 166 | 1.22 |
| 188 | 5 | 194 | 1.3 |
| 293 | 9 | 303 | 1.5 |
| 105 | 4 | 109 | 1.85 |
| 179 | 9 | 187 | 2.43 |
| 173 | 11 | 182 | 3.05 |
| 185 | 16 | 197 | 4.11 |
| 145 | 17 | 157 | 5.48 |
| 174 | 28 | 193 | 7.35 |
| 122 | 27 | 140 | 9.79 |
| 151 | 42 | 179 | 11.94 |
| 121 | 72 | 172 | 21.66 |
| 56 | 54 | 97 | 29.4 |

**Supplementary Table 1. Parameters used for theoretical calculations.** *Parameters n, m and q for our chosen supercells for various misalignment angles θ.*

# Supplementary Notes

## Supplementary Note 1 – Self rotation

Further to the Flake outlined in the main text, here we demonstrate another flake with self-rotation upon annealing. In this case, a fold in the graphene sheet creates a discrete boundary at which the rotation angle changes w.r.t the hBN substrate even prior to annealing. Supp. Fig. 1A shows an optical image of the flake, supp. The left hand side of the flake is aligned with 0.5° to the hBN substrate as shown by Raman data (supp. Fig. 2A) and AFM data ($L$=12.5nm), which can be related to each other and the misalignment angle[1-5]. Fig. 3E shows an AFM topography image of the flake, including the fold. The Right hand side of the flake is aligned with an angle of 1° (with $L$=10nm) to the hBN flake, which is, again, confirmed by Raman spectroscopy (supp. Fig. 2A) and AFM (supp. Fig. 1B, D). It is also partially reconstructed with $\delta/L$=0.4. The Sample has then been annealed to 250°C in a forming gas (Ar 90% and $H_2$ 10%) for 4 hours.

The right hand side of the flake shows considerable change after annealing. Broadening in the Raman spectrum after annealing, displayed in the maps in supp. fig. 2A and C respectively, suggests that the degree of reconstruction within the graphene flake has changed and, interestingly, varies along its length. The moiré periodicity changes, as is shown in supp. Fig. 1D (before) and E (after) to $L$=12.5nm. This is a change of 25% from $L$=10 nm before annealing. The flake is also further reconstructed, with $\delta/L$=0.2, whereas prior to annealing in was $\delta/L$=0.4.

As indicated in the Raman map in supp. Fig. 2C, the flake undergoes a novel self-rotation upon annealing. The further from the fold in the graphene sheet, the more the flake has rotated. It varies from $L$=12.9nm at the very end, to $L$=11.8nm closer to the fold, which is shown in supp. Fig. 3 (A) – (D). Interestingly, all parts of the flake have changed with respect to the position prior to annealing. The most likely explanation for this variation along the length is that the fold in the graphene sheet acts as a barrier to rotation, whose influence decays with distance.

Finally, further to the high resolution AFM and Raman spectroscopy evidence for rotation, supp. Fig 4 displays the large translation of the graphene flake relative to the hBN substrate. A translation of 300 ± 90 nm is measured from before annealing to after annealing.

## Supplementary Note 2 - Flake Homogeneity

Moiré pattern were imaged at various points on the flake (detailed in the main text) to confirm the homogeneity of its period. Supp. Fig. 5A, B, and C show the Young's modulus signal of the moiré pattern at different regions. Unfortunately, some contamination remains on the surface and is visible in (A) and (B). However, there are clean regions, such as (C). The images confirm that the periodicity of the moiré pattern is uniform across the flake.

## Supplementary Note 3 - Wrinkled flakes

Many samples show the characteristic wrinkling upon thermal annealing and cooling. Supp. Fig. 6 gives several examples of very pronounced 1D wrinkling on top of the moiré superlattice for several samples which showed no self-rotation. The wrinkles are always linked to the moiré pattern. Similar, corrugation has been seen recently in the graphene-hBN system[6].

Further, to eliminate the possibility that the observed 'wrinkling' is not a result of noise we have performed several basic studies on how it varies with different scanning parameters. Scan angle, scan size, and scan rate have all been changed independently. The results for one sample are displayed in Supp. Fig. 7. Electronic noise and vibrational noise are the two main causes of apparently 1D patterns in raster scanning studies. Since the 1D patterns absolute period and orientation change appropriately with each of these parameters, we can conclude that the observation is of a real physical phenomenon.

## Supplementary Note 4 - Uniaxial wrinkling

Our model for unidirectional wrinkling (observed in some of our devices) requires development of uniaxial strain, perpendicular to the wrinkles. In order to test this assumption we performed polarised Raman scattering spectroscopy with the incident light polarisation being parallel and perpendicular to the wrinkles, supp. Fig. 8. We indeed observed that the Raman 2D peak for the polarisation perpendicular to the wrinkles is significantly broadened (FWHM 45.0 cm$^{-1}$) in comparison to that with polarisation parallel to the wrinkles (FWHM 40.9 cm$^{-1}$). Note, that no such anisotropic broadening has been observed in the samples which exhibited self-rotation and has not demonstrated wrinkling.

## Supplementary Note 5 - Flakes self-rotation: Evidence from transport

Due to electron interference, moiré superstructure results in reconstruction of the electronic spectrum of graphene. In particular, secondary Dirac points appear in the spectrum, which can be detected either through direct measurements of the density of states in STM[1], or in transport, where additional peaks in resistance can be seen when the Fermi level reaches the secondary Dirac points[2,3,7], supp. Fig. 9. The carrier concentration at which the secondary Dirac points is achieved ($n_{SDP}$) is directly related[3] to the period of the moiré structure as $L=2(\pi/3n_{SDP})^{1/2}$, and could be converted to the misalignment angle[1]. The blue curve in supp. Fig. 9 represents the resistance as a function of the carrier concentration after the initial annealing at 175°C. Converting the position of the resistance peak at the holes side to the misalignment angle we get 0.49°. The red curve was measured on the same sample after an additional annealing at 200°C. The peaks are shifted towards the lower carrier concentration, which converts to the misalignment angle 0.47°. The small (compared to that presented in the main text) rotation of the flake is associated with the fact that the van-der-Waals potential as a function of the misalignment angle saturates below 0.7° and also with the fact that this graphene flake was clamped by contacts, essential for transport measurements.

## Supplementary Note 6 - Self-rotation of graphene on graphene

In order to demonstrate self-rotation of other crystals, beyond the graphene/hBN pair and provide alternative evidences for such rotation, we fabricated $Si/SiO_2$/graphene/hBN/graphene tunnelling device (tunnelling current is being measured between the two graphene electrodes, 3 layers of hBN serve as a tunnelling barrier and silicon substrate is used as a back gate). The tunnelling differential conductivity as a function of the gate $V_g$ and bias $V_b$ voltages for such a device is presented at Supp. Fig. 10A. The low conductivity features, highlighted by red and blue theoretical curves (see [8] for the details of the electrostatic model which has been used to obtain the two lines), originate from the suppression of the tunnelling when the Fermi level in one of the graphene layers passes through the Dirac point, where the density of states is zero. The specific shape of the features represents the linear spectra of graphene (and the fact that the Fermi energy is a square root function of the carrier concentration).

Then, an additional graphene layer was transferred on the top graphene electrode (so the whole structure now is $Si/SiO_2$/graphene/hBN/graphene/graphene). We estimate the misalignment between the two top graphene layers to be approximately 3°. At such misalignment angle the two graphene layers act independently, with the linear dispersion relation still being preserved at low energies in either of them. This fact is confirmed by the observation (Supp. Fig. 10B) of the low-conductivity features (also highlighted by red and blue simulation curves) which are very similar to those observed in Supp. Fig. 10A. Note however, that the green curve in Supp. Fig. 10B is slightly steeper than the blue curve in Supp. Fig. 10A, which reflects the fact that the density of states (still being a linear function of energy) is now doubled for the electrode consisting from two graphene layers.

However, once the sample was annealed at 200°C, the d$I$/d$V_b$ dependence on $V_g$ and $V_b$ becomes strikingly different, Supp. Fig. 10C. Now only one differential conductivity minimum, which looks like a straight line in the coordinate $V_g$ and $V_b$ is visible (highlighted by blue theoretical curve). We interpret this behaviour as the top graphene layers rotating to form Bernal-stacked bilayer. Since bilayer graphene has a parabolic spectra and, as a consequence, constant density of states as a function of energy, it doesn't contribute with any special features in the conductivity. The minimum in d$I$/d$V_b$ is originating from the Fermi level in the bottom graphene passing through the Dirac point. The linear behaviour of this line is characteristic of the constant density of states in the top bilayer graphene electrode (here the same electrostatic model as before[8] was used, just the constant density of states for the top electrode was utilised).

Such behaviour demonstrates that similarly to graphene on hBN system, graphene on graphene is also capable to self-rotate into the energetically favourable aligned configuration.

## Supplementary Note 7 - Modelling methodology

As the ratio between the lattice constants of graphene and hBN is approximately 55/56, a common supercell can be constructed by repeating 56x56 unit cells of graphene on 55x55 unit cells of h-BN, resulting in a 1.8% mismatch in lattice constant. This construction yields an aligned sample (θ=0) with a 1.8% mismatch of the lattice constants without any initial stretching of graphene.

When lattice relaxation is allowed in the energy minimization - a non-uniform pattern of distortion (Supp. Fig. 11) is found. The spread between local compression and expansion observed in our calculations is from 1.391 Å to 1.41 Å with respect to the equilibrium interatomic distance given by the REBO potential as 1.3978 Å. In supp. Fig. 12, we show the dependence on $\theta$ of the maximum and minimum bond-lengths which are almost the same at large angles and show a marked change below $4°$. Here we would like to stress that the average (over the unit cell) stretching/compression of graphene is quite small (well below 1.8%), resulting in only a small lost in the elastic energy, which is compensated by the gain in the van der Waals energy.

To create a common supercell for misaligned graphene on h-BN, we create a supercell which satisfied periodic boundary conditions in the plane for a layer of h-BN rotated by an angle θ with respect to the graphene layer. For some angles, periodic boundary conditions are only satisfied for extremely large supercells. Therefore we allow an adjustment of the lattice constant of graphene of less than 0.01%.

More precisely, we rotate the h-BN by

$$\theta = \frac{1}{2}\arccos\left[\frac{2n^2 + 2nm - m^2}{2(n^2 + nm + m^2)}\right] \quad (1)$$

resulting in a cell of length

$$a_{cell} = a_{BN}\sqrt{n^2 + nm + m^2} \quad (2)$$

Where $a_{BN}$ is the lattice constant of h-BN and *n,m* are positive integers with *n>m*. The graphene layer is not rotated but the unit cell is repeated *q* times, resulting in a side of length

$$a_{cell} = a_{C'}q \quad (3)$$

where *q* is an integer and the lattice constant of graphene $a_{C'}$ is slightly adjusted to make both supercells the same size. By choosing *n, m* and *q* appropriately, stretching is kept to a minimum (<0.01%).

Details of the studied structures are given in Table I.

The results of energy minimization are shown in the main text. Here we show in supp. Fig. 13, the dependence of the out of plane displacements on the misalignment angle *θ*. One can see that displacements start occurring below $\theta \approx 15°$ and saturate around $4°$.

## Supplementary References